# MULTIDIMENSIONAL DATAWAREHOUSE WITH COMBINATION FORMULA

Spits Warnars Harco Leslie Hendric
(BLLES) Budi Luhur Learning Solution - Budi Luhur University
Jl. Petukangan Utara, Kebayoran Lama, Jakarta Selatan 12260, Indonesia
spits@bl.ac.id

**ABSTRACT**

Multidimensional in data warehouse is a compulsion and become the most important for information delivery, without multidimensional datawarehouse is incomplete. Multidimensional give the able to analyze business measurement in many different ways. Multidimensional is also synonymous with online analytical processing (OLAP). By using some concepts in datawarehouse like slice-dice, drill down and roll up will increase the ability of multidimensional datawarehouse. The research question and the discussing for this paper are how much deepest the multidimensional ability from each fact table in datawarehouse. By using the statistic combination formula we try to explore the combination that can be yielded from each dimension in hypercubes, the entire of dimensi combination, minimum combination and maximum combination.

## 1. Introduction

Datawarehouse is not a new hardware or a new software but a computer environment in order to use the database to deliver the strategic information become faster and reliable. Datawarehouse is created by ETL (Extraction Transformation Loading) process that get the data from TPS (Transactional Processing System) or OLTP (OnLine Transactional Processing).

There are 2 kinds of table in datawarehouse first is call fact table and the other is dimensional table. The fact table is the collection of some of foreign key attribute, where each of foreign key become connection to dimensional table and become primary key for each dimensional table. Mostly the primary key for fact table is a composite/compound key from some foreign key. Dimensional table containing clarification detail from each information that result from the analyst the needed of the system. The fact table represent the business dimension of the datawarehouse. Mostly the fact table represent 3 dimension that draw with cube and draw with hypercubes if we represent more than 3 dimension. Picture 1 and picture 2 show up a cube and hypercubes example.

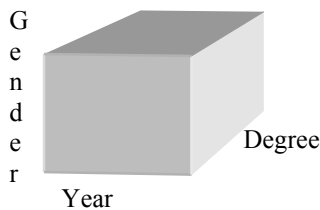

Picture 1 : Cubes

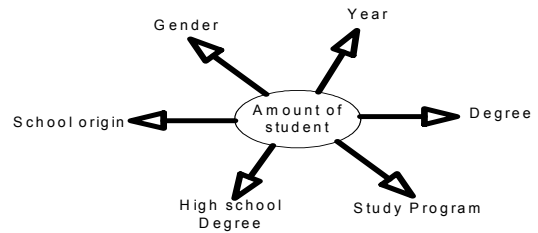

Picture 2 : Hypercubes

From hypercubes or cube will build the logical data model for each of fact table and dimensional table.

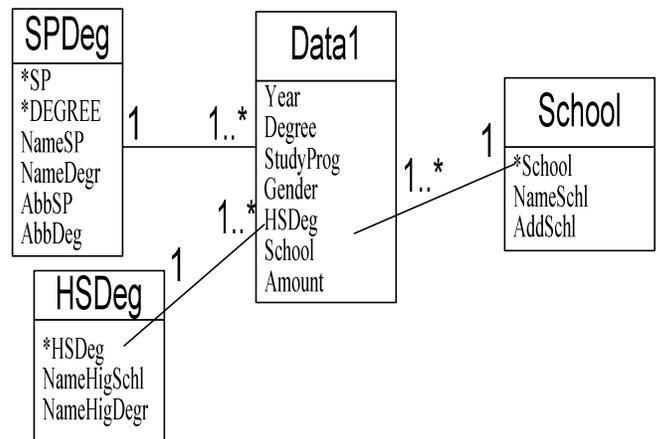

Picture 3: Fact table Data1 and dimensional tabel SPDeg, HSDeg and School

With a simple statement query that access fact table will create a temporary database table that has a valuable if shown in graphical report. Being compare with the number report, the graphical report appearance will increase the readability of data in order to support management make the right decision. Some other people said that picture have so many meaning compare with number. Management will be easy to see the tren degradation or increasing by the graphical report compared with the number report.

For example watch carefully data from the student database fact table below :

| Year | Deg | SP | Gen | Amn |
|------|-----|-----|-----|-----|
| 2000 | 5 | 11 | p | 11 |
| 2000 | 5 | 11 | w | 22 |
| 2000 | 3 | 11 | p | 12 |
| 2000 | 3 | 11 | w | 13 |
| 2000 | 5 | 22 | p | 10 |
| 2001 | 5 | 11 | w | 33 |
| 2001 | 5 | 11 | p | 44 |
| 2001 | 3 | 11 | w | 14 |
| 2001 | 3 | 11 | p | 15 |
| 2002 | 5 | 11 | p | 55 |
| 2002 | 5 | 11 | w | 66 |
| 2002 | 3 | 11 | p | 16 |
| 2002 | 3 | 11 | w | 17 |

Table 1 : Student database fact table

From that example data we given the sql statement below :
  select year, Deg,SP,Gen,sum(Amn) as Amount
    from dwmhs where year="2000"
      group by Year, Deg,Gen;
From the sql statement up there will result the report :

Year : 2000

| Gen | S1 | D3 |
|-----|-----|-----|
| Men | 21 | 12 |
| Women | 22 | 13 |

Table 2 : Number report

The report on table 2 will be shown in graphical report like picture 4 down here

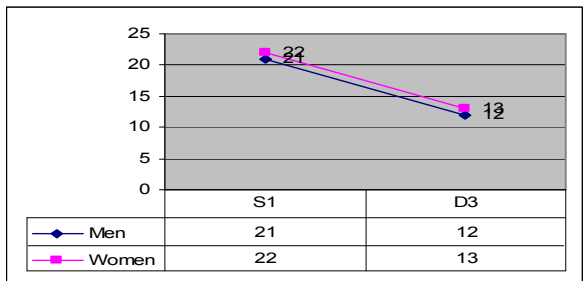

Picture 4 : Graphical report

From the difference between number report on table 2 and graphical report on picture 4, we can see how the management will be helped dan satisfy when they look the report that shown the tren degradation or increasing. The management will get headache when they see the number report like table 2, even sometimes they need some reports like number report.

**Slice dan Dice**

The ability multidimensional datawarehouse that build from hypercubes or cubes in order to give the better and faster information for decision support can increase with slice and dice concept. Slice and dice that one of the multidimensional datawarehouse concept where hypercubes or cube can view from many dimension. Beside that slice and dice have a function to take a pieces of hypercubes or cube based on to certain value in one or some dimension. Slice and dice can be done by giving the simple sql statement.

Example Picture 5 below describe how we can slice and dice process to the data just like the management desire. The management can view the data from many dimension and even from the combination some of dimension. From the way to view the data from many combination the management can find so many possibility decision, even they can find some data mining there.

Year : 2000

| Gender | S1 | D3 |
|--------|-----|-----|
| Men | | |
| women | | |

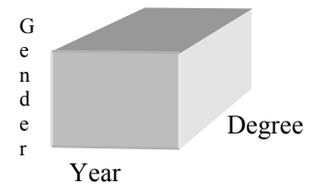
Degree
Year

Degree: Bachelor

| Year | M | W |
|------|---|---|
| 2001 | | |
| 2002 | | |
| 2003 | | |

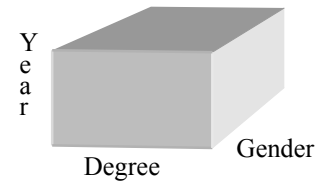
Degree
Gender

Gender : Men

| Gender | 2001 | 2002 |
|--------|------|------|
| S1 | | |
| D3 | | |

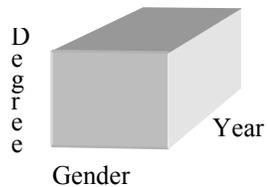
Year
Gender

Picture 5 : Slice and Dice concept

**Roll up dan Drill Down**

Beside use slice and dice, the multidimensional ability of hypercubes or cube in order to deliver the better and faster information for getting decision will be increased by using roll up and drill down concept. Roll up is a generalization process one or more dimension with embrace and summarize the value of size measure. With other word roll up is generalization that have meaning ascend one up level in dimension hierarchy and drill down process is the process to choose and show detail data in one or some dimension and of course the opposite of roll up process.

The same with slice and dice concept can be done by using a simple sql statement, the roll up and drill down concept can also be done by using a simple sql statement. To make understand the explanation let see the picture 6 down here where the roll up concept can be done by a sql statement :

"Select year, sum(amn) as amount from DWmhs group by year "

The amount 68 is the sum up from 12,13,21 and 22
The amount 106 is the sum up from 15,14,44 and 33
The amount 154 is the sum up from 16,17,55 and 66

| Year | Deg | Gend | Amn |
|------|-----|------|-----|
| 2000 | 3   | m    | 12  |
| 2000 | 3   | w    | 13  |
| 2000 | 5   | m    | 21  |
| 2000 | 5   | w    | 22  |
| 2001 | 3   | m    | 15  |
| 2001 | 3   | w    | 14  |
| 2001 | 5   | m    | 44  |
| 2001 | 5   | w    | 33  |
| 2002 | 3   | m    | 16  |
| 2002 | 3   | w    | 17  |
| 2002 | 5   | m    | 55  |
| 2002 | 5   | w    | 66  |

| Year | Amn |
|------|-----|
| 2000 | 68  |
| 2001 | 106 |
| 2002 | 154 |

Picture 6 : Roll up concept

Opposite from picture 6 up there, the picture 7 down here will explanation about drill down concept.

| Year | Deg | SP | Gend | Amn |
|------|-----|----|------|-----|
| 2000 | 5   | 11 | m    | 11  |
| 2000 | 5   | 11 | w    | 22  |
| 2000 | 3   | 11 | m    | 12  |
| 2000 | 3   | 11 | w    | 13  |
| 2000 | 5   | 22 | m    | 10  |
| 2001 | 5   | 11 | w    | 33  |
| 2001 | 5   | 11 | m    | 44  |
| 2001 | 3   | 11 | w    | 14  |
| 2001 | 3   | 11 | m    | 15  |

| Nim        | Name    | Gend |
|------------|---------|------|
| 0011500001 | Joni    | m    |
| 0011500002 | Tono    | m    |
| 0011500005 | Edi saku| m    |
| 0011500007 | Feri    | m    |
| 0011500010 | Bono    | m    |
| 0011500012 | Dirun   | m    |
| 0011500013 | Gunawan | m    |
| 0011500014 | Hari    | m    |
| 0011500021 | Tomi    | m    |
| 0011500028 | Budi    | m    |
| 0011500030 | Lukas   | m    |

Picture 7 :Drill down concept

Picture 7 up there will explanation how drill down concept will show the detail data from a record that circle on the first row. That drill down concept can be done by a simple sql statement :

Select a.nim,a.nama from mastmhs a, DWmhs b
  where left(a.nim,2)=right(b.ang,2) and
        substr(a.nim,3,2)=b.ps and
        substr(a.nim,5,1)b.jenj and a.jenkel=b.jenkel

The decision support manajemen will be helped in order to make a decision by view a hypercubes or cube with roll up generalization and detail drill down. The generalization data with roll up concept will help the management in order to make a decision with a data summary and the detail data with drill down concept will help the management in order to make a decision with detail data.

**The research question**

By description up there we can see that the multidimensional concept will be implemented by slice-dice, roll up and drill down concept in order to increase the ability of hypercubes or cube to give the decision support information for decision support management.

Increasing the ability hypercubes or cube to give the decision support information will generate the research question that will be discussed in the next explanation. Next some research question like :
1. How many minimal dimension and maximal dimension that can be created from a hypercubes or cube ?
2. How many combination report or graphic that can be created from a hypercubes or cube ?
3. How many combination report or graphic that can be created on each dimension from a hypercubes or cube ?

**Graphical report**

The graphical report is an intersection between horizontal line and vertical line. The horizontal line on the graphical report can only draw one value of column while vertical line draw one value of column or combination more than one column. Column can have meaning become dimension on hypercubes or cube.

The maximal column that can be used become horizontal line is amount column on fact table or amount dimension on hypercubes or cube beside amount column. Amount column can't be acted become a vertical line or horizontal line because amount column is a value that will be showed on the intersection vertical line and horizontal line on graphical report. This amount column will show the the tren degradation or increasing value on the graphical report.

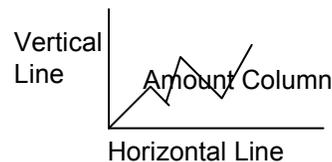

Picture 8 : Intersection vertical line and horizontal line

Previously we know that vertical line is a dimension that can be created from one column/dimension or combination value of column/dimension. By definition up there we know that maximal dimension from fact table is an amount fact table column beside amount column. If a table have 2 column beside amount column will have maximal 2 dimension, if having 3 column beside amount column will have maximum 3 dimension and so on.

These are can be viewed on hypercubes or cube where the amount dimension on hypercubes or cube is the same with amount column on fact table beside amount column. Watch the picture 9 down here where the amount dimension on hypercubes are 6 where the same with amount field on fact table data1 beside field Amount are 6 too.

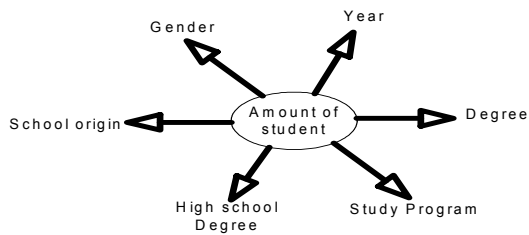

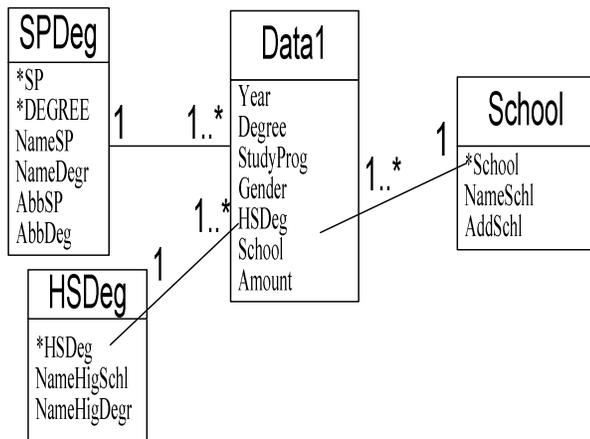

Picture 9 : Hypercubes and fact table Data1

Thereby the research question about how many minimal dimension and maximal dimension that can be created from a hypercubes or cube already be answered where amount minimal dimension is 1 and amount maximal dimension as many as dimension that owned by hypercubes or cube.

**Combination Formula**

To get the report/table combination that can be created from each dimension on hypercubes without pay attention the sequence of formation and also to answer the research question about how many report or graphics combination that can be created from a fact table hence it is better we use combination formula on statistic science.

Combination theory said that : "Combination from a number of n different object that take a number r at one moment is the choosing r object without paying attention to its formation sequence."(Jong Jek Siang, 2002).

The amount of object combination is amount of n that get r at one moment having formula as follows
$n\,C\,r$ or $C(n,r)$ or $C\,n,r$ or $\binom{n}{r}$

where : $n\,C\,r = \dfrac{n!}{r!\,(n-r)!}$

$n! = n(n-1)(n-2)\ldots 1$

So :   $0! = 1$
       $1! = 1$
       $2! = 2*1$         $= 2$
       $3! = 3*2*1$       $= 6$
       $6! = 6*5*4*3*2*1$ $= 720$

By using combination formula up there hence deployed a formula that can be show the amount report or graphic that can be created based on the dimension amount on hypercubes to fulfill multidimensional concept on datawarehouse. The deployed formula is :
$n * {}_{n-1}C_{r-1}$
where :
  n is amount of hypercubes dimension
  r is a dimensional value that will be created

By deploying this formula we will know
1. The overall of report/graphic combination that can be created
2. The report/graphic combination on each dimension

Next will arise the research question :
Why the discussing in order to get dimension on hypercubes or cube deployed from theory statistic combination formula and why not use the other formula theory. For answering that question it is better that we discuss some reason why we choose theory statistic combination formula.

For making the report more than 2 dimension, the first dimension/column have the static characteristic while the next dimension/column beside amount column can be changed their position. This first dimension/column can't be changed because the first dimension/column becoming value of horizontal line on graphic, while the amount column is the value that will be showed on the intersection between row and column on graphic view.

For example look at table 3 down here

| Year | Degree | Gender | Amount |
|---|---|---|---|
| 2000 | 5 | M | 21 |
| 2000 | 5 | W | 22 |
| 2000 | 3 | M | 12 |
| 2000 | 3 | W | 13 |
| 2001 | 5 | M | 44 |
| 2001 | 5 | W | 33 |
| 2001 | 3 | M | 15 |
| 2001 | 3 | W | 14 |
| 2002 | 5 | M | 55 |
| 2002 | 5 | W | 66 |
| 2002 | 3 | M | 16 |
| 2002 | 3 | W | 17 |
| Total | | | **328** |

Table 3 : report

If table 3 up there transform into graphical will have looking view graphic like picture 10 down here.

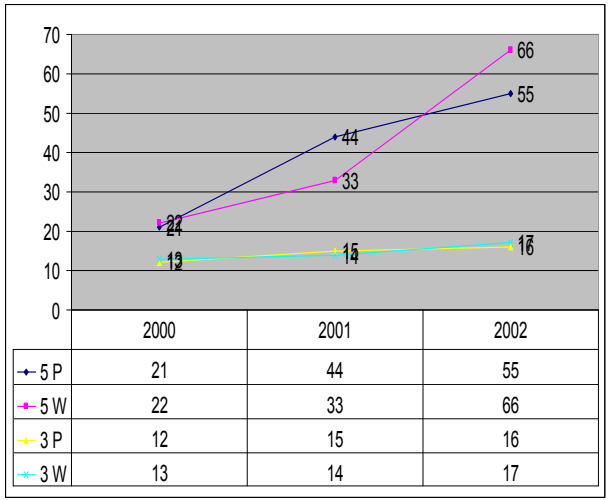

Picture 10: graphical report

The report/table view on table 3 up there can have the report view that can change the column position beside the first column and amount column.

| Year | Gender | Degree | Amount |
|---|---|---|---|
| 2000 | M | 5 | 21 |
| 2000 | W | 5 | 22 |
| 2000 | M | 3 | 12 |
| 2000 | W | 3 | 13 |
| 2001 | M | 5 | 44 |
| 2001 | W | 5 | 33 |
| 2001 | M | 3 | 15 |
| 2001 | W | 3 | 14 |
| 2002 | M | 5 | 55 |
| 2002 | W | 5 | 66 |
| 2002 | M | 3 | 16 |
| 2002 | W | 3 | 17 |
| Total | | | **328** |

Table 4 : report witch switch column 2 and 3

On table 4 up there the position column 2 switch with column 3. If table 4 transform into graphical will have looking view graphic like picture 11 down here.

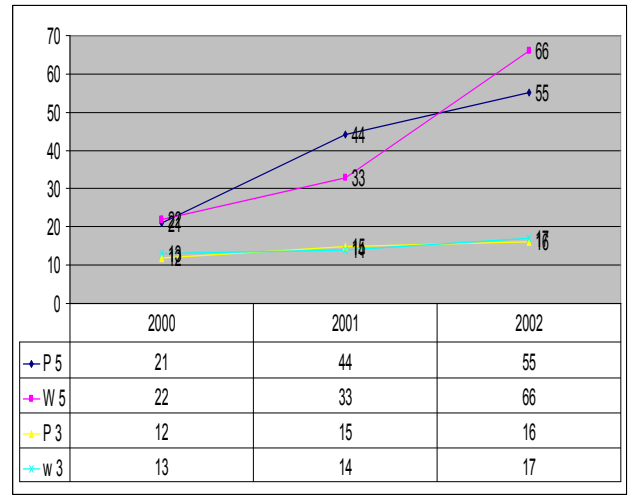

Picture 11: graphical report switch column 2 and 3

Looking to table 3 and 4 we can see that there is no different for comparing between table 3 and table 4, the value on column amount have the same value. The differentiating only switching between column gender and degree.

And so do with the appearance both two of graphic up there seen have the same appearance, having horizontal line contain year that already the same. The differentiating only view on the appearance graphic legend where line blue legend color on picture 10 is 5p while on picture 11 is P5.

From the explanation comparing between two report up there we can see that dimension combination to be searched is combine a number of ***n*** different object that get from a number ***r*** without pay attention to formation sequence. These explanation in line with contain and explanation combination formula theory, so by that in order to get dimension on hypercubes or cube will be used statistic formula combination theory.

**Formula combination 3 dimension**

To proof the development combination formula up there, we have to trial-error and test the development combination formula by taking data sample from a hypercubes or cube 3 dimension

To make easy the explanation every each dimensional will be represented by alphabetic letter so that the hypercubes or cube will have the report appearance like table 5 down here.

| A | B | C | Jumlah |
|---|---|---|---|
|   |   |   |   |
|   |   |   |   |

Table 5: the report dimensi A,B,C

The hypercubes for table 5 up there will be like this

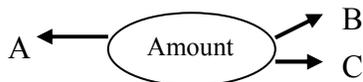

Picture 12 : hypercubes 3 dimension (A,B,C)
By using combination formula :    $n * n{-}1 C r{-}1$
Where n is an amount of hypercubes dimensi and r is dimensi value that will be created. Because the amount hypercubes dimensi is 3 then n=3. According to data warehouse multidimensional concept, hypercubes 3 dimensi will be formulate the combination from report/table or graphic that can be created on each dimension.

| Dimensi 1, meaning r=1 | Dimensi 2, meaning r=2 |
|---|---|
| then = n * n-1 C r-1 | then = n * n-1 C r-1 |
| = 3 * 3-1 C 1-1 | = 3 * 3-1 C 2-1 |
| = 3 * 2  C 0 | = 3 * 2  C 1 |
| = 3 * ( n! ) | = 3 * ( n! ) |
|     r! (n-r) ! |     r! (n-r) ! |
| = 3 * ( 2! ) | = 3 * ( 2! ) |
|     0! (2-0) ! |     1! (2-1) ! |
| = 3 * ( 2! ) | = 3 * ( 2! ) |
|     0! (2!) |     1! (1!) |
| = 3 * ( 1*2 ) | = 3 * ( 1*2 ) |
|     1 (1*2) |     1 (1) |
| = 3 * 1 | = 3 * 2 |
| = 3 | = 6 |

Dimensi 3, meaning r=3
   then  = n * n-1 C r-1
        = 3 * 3-1 C 3-1
        = 3 * 2  C 2
        = 3 * ( n! )
             r! (n-r) !
        = 3 * ( 2! )
             2! (2-2) !
        = 3 * ( 2! )
             2! (0!)
        = 3 * ( 1*2 )
             1*2 (1)
        = 3 * 1
        = 3

The result from combination formula on hypercubes 3 dimension up there can be view on the table below here:

| Horizontal Dimensional | Vertical dimensional | | |
|---|---|---|---|
|  | 1 | 2 | 3 |
| A | A | AB and AC | A BC or A CB |
| B | B | BA and BC | B AC or A CA |
| C | C | CA and CB | C AB or A BA |
| Combination | 3 | 6 | 3 |
| Total : 12 Combination | | | |

$$\sum_{n}^{1} (n*n{-}1 C r{-}1) / n \qquad n*n{-}1 C r{-}1$$

Table 6 : combination 3 dimensional

Based on table 6 up there we can see that formula $n*n{-}1 C r{-}1$ using to count combination amount by vertical dimensional dan using formula :

$$\sum_{n}^{1} (n*n{-}1 C r{-}1) / n$$

To count combination amount by horizontal dimensional.

The intersection between 2 dimensi below here will show there are 6 column intersection are AB, AC, BA, BC, CA and CB.

|   | A | B | C |
|---|---|---|---|
| A |   | AB | AC |
| B | BA |   | BC |
| C | CA | CB |   |

This is according with verification result with formula when dimensi 2 and will yield 6 combination report/table/grafik.

The intersection between 3 dimension below here will show there are 6 column intersection are A BC, A CB, B AC, B CA, C AB dan C BA.

|   | A | | B | | C | |
|---|---|---|---|---|---|---|
|   | B | C | A | C | A | B |
| A |   |   |   | A BC |   | A CB |
| B |   | B AC |   |   | B CA |   |
| C | C AB |   | C BA |   |   |   |

According to previous explanation that for making report more than 2 dimensional, the first column is stable while the next column beside amount column can change their position. This first column become horizontal line on graphic appearance. Therefore the 6 intersection column up there cause have the same report appearance and graphic appearance can be switch the next column beside first column and amount column.

Thus because of 6 intersection column up there having equal first column and the next column have the same appearance that can be switched their position are
A  BC    or  A  CB
B  AC    or  B  CA

C AB   or C BA

In fact there are 3 column intersection and as according to verification result with formula at the time dimensi 3 will produce 3 combination report/table/grafik. Seen by verification combination formula on hypercubes 3 dimension, the first combination and the end will having the same value with dimensional value is 3.

**Conclusion**

Finally we will get some conclusion are :
1. The value for first and end combination will have the same value with dimensional amount beside amount column on fact table or hypercubes.
2. The combination formula can making easy and become reference to make an OLAP application (Online Analytical Processing) that can access data warehouse and can show the multidimensional ability from a hypercubes or cube in more maximal.
3. Conceptly sql statement that using to access data warehouse hypercubes having the same like this :
   select field1..fieldn, sum(amn) as amount
       from tablename
           group by field1..fieldn;
   where the sequence select field1…fieldn
   will be the same with group by field1…fieldn

**Full Author Information**

Spits Warnars Harco Leslie Hendric was born on 21 April 1972. Start get the bachelor degree in 1991 on STMIK Budi Luhur with tenure scholarship and finish in 1995. Become a lecturer in 1995 until right now on Faculty of information technology Budi Luhur university. Beside that from 1995-1996 become a programmer in Data Processing Center STMIK Budi Luhur. In 1996 get a promote become an analyst system on STMIK Budi Luhur till 2002. From 2002 till June 2006 become a secretary of information system study program, faculty of information technology Budi Luhur university. From June 2006 till right now become vice of director BLESS (Budi Luhur Learning Solution).

In 2004 Continue computer master degree at study program of Information Technology at faculty of Computer Science University of Indonesia. Right now still finishing tesis for master degree with title "Data Warehouse design in high education. Study case faculty of information technology Budi Luhur University". My Personal Web on WWW.SPITS.8K.COm or my Update Cv on www.bl.ac.id/dosen/spits/myCVJune2006.htm.